\documentclass[aps,prl,preprint,showpacs,nofootinbib]{revtex4-1}
\usepackage{graphicx}

\begin{document}


\title{Low-Temperature Light Detectors: Neganov-Luke Amplification and Calibration}

\author{C. Isaila$^{1,2}$, C. Ciemniak$^{1}$, F. v. Feilitzsch$^{1}$, A. G\"utlein$^1$, J. Kemmer$^3$, T. Lachenmaier$^{1,2,4}$, J.-C. Lanfranchi$^{1,2}$, S. Pfister$^1$, W. Potzel$^1$, S.~Roth$^{1,2}$, M. v. Sivers$^1$, R. Strauss$^1$, W. Westphal$^1$, and F. Wiest$^3$}

\affiliation{$^1$Physik-Department E15, Technische Universit\"at M\"unchen, 85748 Garching, Germany\\
$^2$Excellence Cluster "Universe", Technische Universit\"at M\"unchen, 85748 Garching, Germany\\
$^3$KETEK GmbH, Hofer Strasse 3, 81737 M\"unchen, Germany\\
$^4$Physikalisches Institut, Eberhard-Karls-Universit\"at T\"ubingen, 72076 T\"ubingen, Germany}








\date{\today}

\begin{abstract}
The simultaneous measurement of phonons and scintillation light induced by incident particles in a scintillating crystal such as CaWO$_4$ is a powerful technique for the active rejection
of background induced by $\gamma$'s and $\beta$'s and even neutrons in direct Dark Matter searches. However, $\lesssim1\%$ of the energy deposited in a CaWO$_4$ crystal is detected as light. Thus, very sensitive light detectors are needed for an efficient event-by-event background discrimination. Due to the Neganov-Luke effect, the threshold of low-temperature light detectors based on semiconducting substrates can be improved significantly by drifting the photon-induced electron-hole pairs in an applied electric field. We present measurements with low-temperature light detectors based on this amplification mechanism. The Neganov-Luke effect makes it possible to improve the signal-to-noise ratio of our light detectors by a factor of $\sim$9 corresponding to an energy threshold of $\sim$21 eV. We also describe a method for an absolute energy calibration using a light-emitting diode.
\end{abstract}

\pacs{29.40.Mc, 63.20.-e, 72.20.Jv, 74.78.-w, 95.35.+d}

\maketitle

One of the main objectives of contemporary astroparticle physics is solving the Dark Matter enigma. Among several hypothetical particles that might account for Dark Matter, WIMPs (Weakly Interacting Massive Particles) play a central role, see, e.g., \cite{Taoso}.
The aim of the CRESST (Cryogenic Rare Event Search with Superconducting Thermometers) experiment is the direct detection of WIMPs via coherent elastic scattering off the nuclei in a terrestrial target \cite{cresst},\cite{cresst1}. Such an interaction causes a recoil of the nuclei. The recoil energy is transformed into phonons and - to a much lesser extent - into light. Background radiation, mainly $\gamma$'s and $\beta$'s, interacting electromagnetically induce electron recoils, the energy of which is also transformed into phonons and light. Compared to nuclear recoils of the same energy the light yield is much larger \cite{cresst},\cite{cresst1}. Due to the different light yields a very efficient discrimination of the background induced by $\gamma$'s and $\beta$'s is achieved.The detection scheme employed in the second phase of CRESST and possibly also in the future EURECA (European Underground Rare Event Calorimeter Array) experiment \cite{EURECA} is based on low-temperature detectors using the phonon-light technique, i.e., the simultaneous measurement of the phonons and the scintillation light induced by incident particles in a CaWO$_4$ target single crystal operated at mK temperatures. The phonons are detected by a superconducting transition edge sensor (TES) \cite{Proebst} on the CaWO$_4$ crystal. The scintillation light is measured by a separate low-temperature light detector based on high-purity silicon or silicon-on-sapphire (SOS) substrates, also equipped with a TES, which measures the phonons generated in the substrate by the photons.

The fraction of the deposited energy in a CaWO$_4$ crystal due to electron recoils detected as light is at the $1\%$ level and this fraction is further reduced for nuclear recoils by the so-called quenching factor (QF) with QF$\gtrsim$10 \cite{cresst1},\cite{Bavykina}. Due to the small number of photons absorbed in the light detector, for nuclear recoils the energy threshold and resolution of the light channel are dominated by electronic noise. To improve the threshold as well as the resolution of the light channel, more sensitive light detectors are needed.

Following Neganov and Trofimov \cite{neganov} and Luke \cite{luke}, the energy threshold of a low-temperature light detector employing a semiconducting substrate can be improved by drifting the photon-induced electron-hole pairs by an applied electric field. Due to the heat dissipated in the substrate by the drifting electron-hole pairs the phonon signal is amplified. If the generated charge is completely collected, the resulting thermal gain $G_t$ is given by
\begin{equation}\label{Neganov-Luke}
G_t=1+\frac{eV}{\epsilon}
\end{equation} 
\noindent where $V$ denotes the bias (Neganov-Luke) voltage, $e$ the electron charge and $\epsilon$ the energy needed to create an electron-hole pair. Detectors exploiting this heat-amplification technique have been investigated for applications in neutrino physics \cite{Akerib}. A detector of the Neganov-Luke type for 122 keV $\gamma$-rays of $^{57}$Co has been described in \cite{Spooner}: Ohmic contacts were set up on two opposite faces of a Si crystal. These were covered by Al layers on both faces. For visible light, however, two main problems arise: The Al layer reflects most of the visible light and, even more important, the light is absorbed within a thin surface layer. A low-threshold cryogenic light detector with Neganov-Luke amplification has been described in Ref. \cite{Stark}. However, these first results showed that the performance of such a cryogenic light detector was seriously impeded by excess noise when a Neganov-Luke voltage was applied.  Because of surface defects and traps, drifting electron-hole pairs imposes a challenging problem which we have tried to solve.

\begin{figure}[htbp]
\begin{center}
\includegraphics[width=0.36\textwidth]{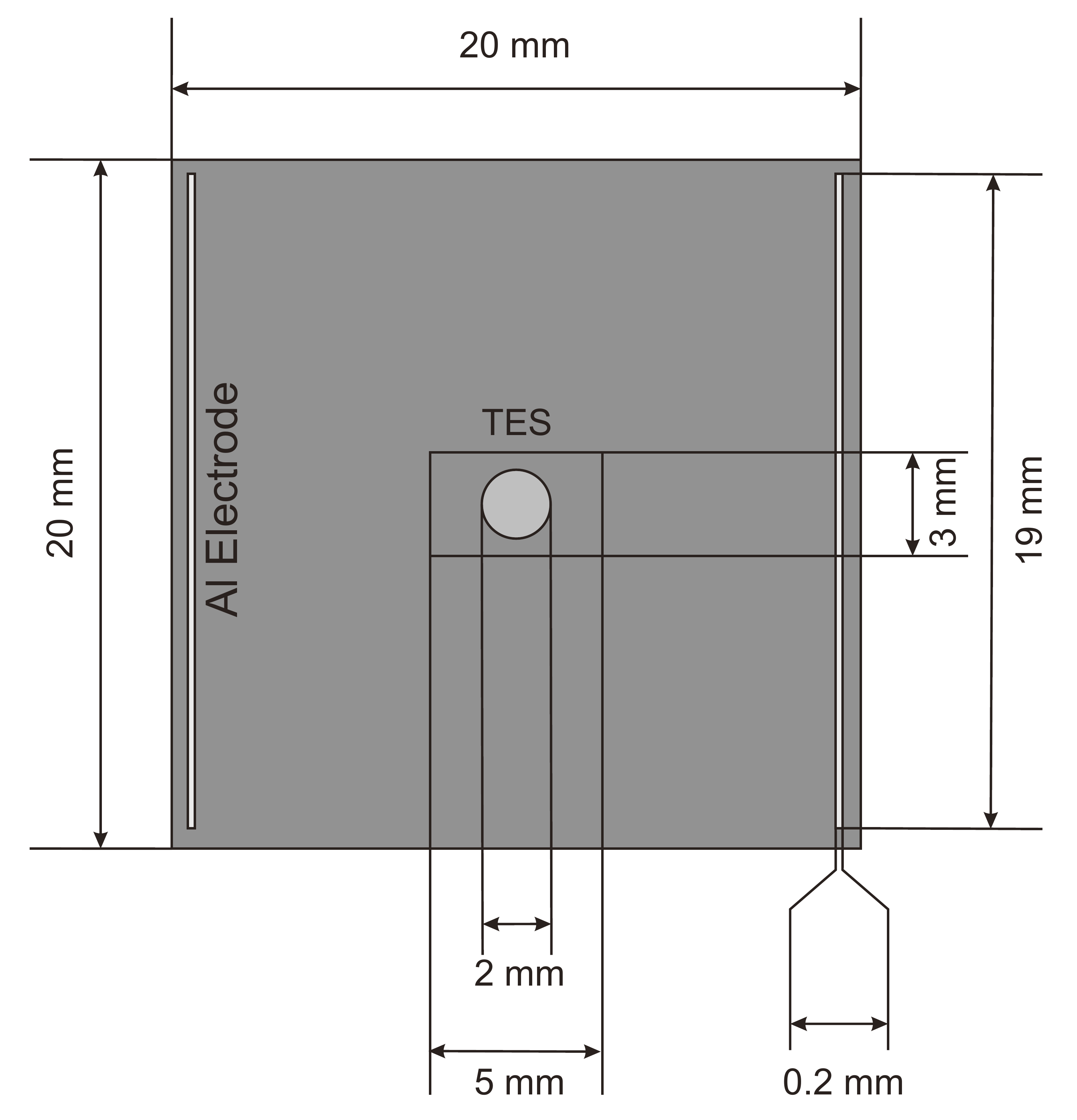}
\caption{\small{Schematic view of the frontside of the composite light detector. It consists of a 20$\times$20$\times$0.5 mm$^3$ silicon absorber and a small (3$\times$5$\times$0.5 mm$^3$) Si substrate carrying the TES. The small substrate is coupled to the absorber by gluing. The absorber is equipped with two (19$\times$0.2) mm$^2$ aluminum electrodes for the application of the Neganov-Luke voltage. The backside of the absorber is exposed to the light source.}}
\label{setup}
\end{center}
\end{figure}

The low-temperature light detectors developed here \cite{Isaila},\cite{IsailaNIM} are based on the composite detector design (CDD) \cite{Roth}. A schematic view of the setup is depicted in Fig. \ref{setup}. The TES is evaporated onto a small (3$\times$5$\times$0.5 mm$^3$) Si disk which is then glued to a 20$\times$20$\times$0.5 mm$^3$ Si plate acting as light absorber. The room-temperature resistivity of the Si is $>$ 10 k$\Omega$cm. For the application of the Neganov-Luke voltage the absorber is equipped with two Al electrodes (19$\times$0.2 mm$^2$ each, separated by $\sim$17 mm) directly evaporated onto the absorber.

The TES used here consists of an Ir/Au bilayer \cite{Nagel} exhibiting a superconducting transition temperature of $\sim$30 mK. A TES is operated within the narrow transition region between the normal and the superconducting state. In this way, the resistance of the film becomes highly dependent on temperature, such that particle interactions that induce a temperature rise of the film also increase the film resistance which is measured via a current change picked up by a SQUID (Superconducting Quantum Interference Device). As compared to the Neganov-Luke detector used in Ref. \cite{Stark}, the new design offers several advantages: It provides almost unhindered access of the light to a large surface, since the backside of the absorber is exposed to the light source. The drifting of the electron-hole pairs occurs through the bulk, not along the surface. In addition, the CDD insures that no Au atoms of the TES can diffuse into the Si absorber and create traps and that the electronic SQUID read-out is electrically decoupled from the Neganov-Luke voltage \cite{Isaila}. 

In this Letter we want to show two aspects: I) The energy calibration of our light detectors can be accomplished by pulses from a light-emitting diode (LED); II) The Neganov-Luke effect provides an amplification of the light signal and leads to an improvement of the energy threshold as well as the energy resolution. In particular, we demonstrate that the excess noise observed in Ref. \cite{Stark} can be avoided to a large extent.

\begin{figure}[ht]
\begin{center}
\includegraphics[width=0.7\textwidth]{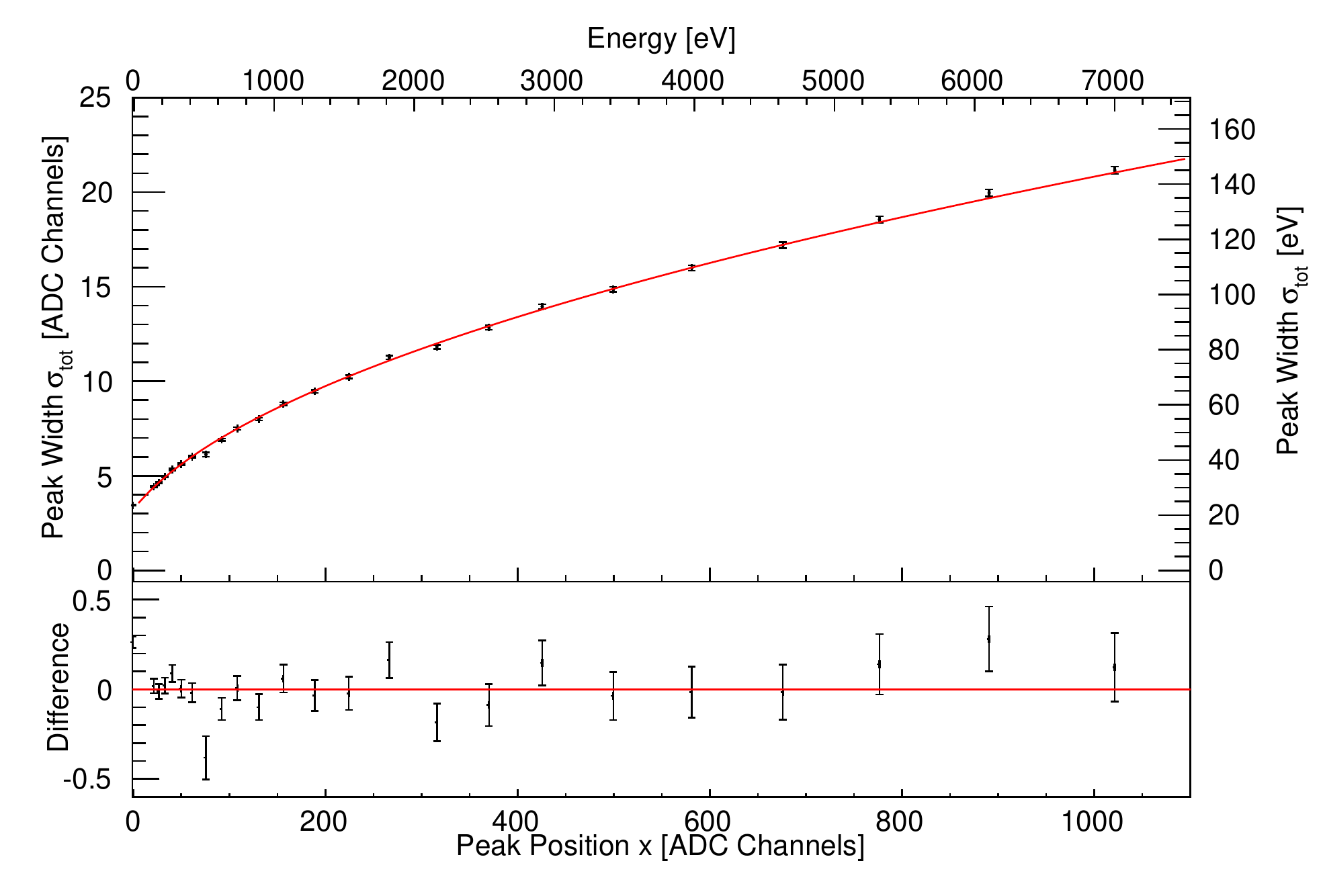}
\caption{\small{Top: Measured peak width $\sigma_{tot}$ of the light peaks versus peak position $x$ fitted by the function $\sigma_{tot}^2 = \sigma_{ph}^2+ax$. The error bars indicate the 1$\sigma$ statistical error. The scale at the top (right) gives the peak positions (widths) in energy units using a photon energy of 2.9 eV emitted by the LED to convert $x$ and $\sigma_{tot}$ into energy. Bottom: Difference (in ADC Channels) between experimental and fitted results. No systematic deviation can be noticed.}}
\label{xsi}
\end{center}
\end{figure}

I) \textit{Calibration.} For the calibration the following scheme was adopted: Light pulses with a length of 500 ns were generated by an InGaN LED\footnote{The LED was operated at room temperature. The light pulses were guided to the low-temperature (mK) region by fiber optic. No significant ($\ll$1mK) temperature rise was observed when the LED pulses were switched on.} and the resulting pulse-height spectra of the detector for a set of light intensities were recorded. The wavelength ($\lambda\approx$ 430 nm corresponding to a photon energy of $\sim$2.9 eV) emitted by the LED matches the spectral output ($\lambda_{\rm{CaWO_4}}\approx$ 443 nm at T=8 K) of the CaWO$_4$ crystal \cite{Mikhailik}. The light peaks were fitted by Gaussian distributions. In the following, the energy of a given light peak is inferred from the peak width of the Gaussian and the corresponding peak position. The total observed width $\sigma_{tot}$ of a light peak induced in a low-temperature light detector depends on several parameters, in particular on electronic noise $\sigma_{el}$, position dependence $\sigma_{pos}$, charge trapping $\sigma_{tr}$, photon statistics $\sigma_{ph}$, and possibly unknown other contributions. We assume $\sigma_{el}$, $\sigma_{tr}$, and $\sigma_{pos}$ to be independent of energy and summarize these three contributions by the constant $\sigma_0$. In addition, we assume $\sigma_{ph}$ and $\sigma_0$ to be independent of each other and to follow Gaussian distributions.
Thus, for the investigated detector the total width $\sigma_{tot}^2$ can be written as:
\begin{equation}
\sigma_{tot}^2 = \sigma_{ph}^2+\sigma_0^2.
\end{equation}
\noindent 
Here, the photon statistics is being considered as Gaussian due to the large number of photons associated with a typical calibration light pulse. Assuming a linear response to the energy deposited by $N$ photons, the measured peak position $x$ (derived from the pulse height) and the corresponding peak width $\sigma_{ph}$ (photon statistics) scale as
\begin{eqnarray}\label{xaN}
x&=&aN\\
\sigma_{ph}&=&a\sigma_N=a\sqrt{N}=\sqrt{ax},
\end{eqnarray}
\noindent where $a$ denotes the scaling factor and $\sigma_N$ the standard deviation of the fluctuating number of absorbed photons which equals $\sqrt{N}$ according to Poisson counting statistics. The relationship between the measured peak position $x$ and the peak width $\sigma_{tot}$ can now be written as
\begin{equation}\label{as0}
\sigma_{tot}=\sqrt{\sigma_0^2+a^2\sigma_N^2}=\sqrt{\sigma_0^2+a^2N}=\sqrt{\sigma_0^2+ax}.
\end{equation}
\noindent We applied this calibration procedure to our light detector. The parameters $\sigma_0$ and $a$ have been derived by fitting $\sigma_{tot}^2 = \sigma_{ph}^2+ax$ to a measured ($x$,$\sigma_{tot}$) set obtained by varying the intensity of the light pulses provided by the LED. This fit is shown in Fig. \ref{xsi} together with the measured widths $\sigma_{tot}$ of the light peaks versus their positions $x$.

The fitted values for the parameters $a$ and $\sigma_0$ are \cite{Isaila}
\begin{eqnarray}\label{a}
a&=&0.423\pm 0.002\\
\label{sigma0}
\sigma_0&=&3.21\pm 0.03
\end{eqnarray}
where the units for a and $\sigma_{0}$ are given in analog-to-digital converter (ADC) channels of the SQUID response.

The pulse heights of the light detector are determined by standard event fits\footnote{In the standard event fit procedure a pulse template (standard event) is obtained by averaging a set of normalized pulses of similar energy. This standard event is fitted to each pulse with three free parameters: baseline offset of the pulse, trigger time offset, and amplitude, the latter being a measure of the energy of the pulse.} which can also be used to estimate the energy threshold $E_{th}$ of the detector from randomly acquired noise samples which result in a peak at zero energy. This peak can be fitted by a Gaussian distribution. For the characterization of our detector, as a figure of merit we define $E_{th}$ as the 5$\sigma$ width of this peak. Using the scaling factor of eq. (\ref{a}) together with eq. (\ref{xaN}), this 5$\sigma$ value can be related to a photon number $N_{th}$. With a photon energy of the InGaN LED of $\sim$2.9 eV we derive from the noise samples an energy threshold of $E_{th}$=$N_{th} \cdot$2.9 eV=118$\pm$1 eV for this detector.

From the LED-calibration data we can also obtain a value for $E_{th}$: Using eqs. (\ref{a}) and (\ref{xaN}) together with the value 5$\sigma_0$ (see eq. (\ref{sigma0})) we get $E_{th}$=110$\pm$2 eV.

In addition, a $^{55}$Fe source was used to check this calibration method. The result obtained from the 5.9 keV $^{55}$Mn$_{K_{\alpha}}$-line is $E_{th}$=119$\pm$2 eV. All three derived energy thresholds are in reasonable agreement, thus validating the above-mentioned calibration method and assumptions for such a type of light detector.

In the phonon-light detection scheme, a low-tem\-perature light detector is always operated in coincidence with a phonon signal generated at the same time as the light signal. This phonon signal usually provides the trigger information. Therefore, the 5$\sigma$ width of the noise peak obtained for light detectors in off-line analysis can indeed be considered as their relevant energy threshold.
  
From the values obtained for $a$ and $\sigma_0$ it becomes evident that the peak width recorded with the light detector is dominated by photon statistics at energies in the keV regime, i.e., $ax\gg \sigma_0^2$. Both contributions, $\sigma_0$ and $\sigma_{ph}$, to the measured peak width are equal for a photon number of $N$$\sim$58. With a photon energy of 2.9 eV this corresponds to an energy of $N\cdot 2.9\approx$170 eV. Below this energy, the peak width in the light detector is dominated by $\sigma_0$, which in turn is mainly determined by electronic and magnetic noise. Since the relevant nuclear recoil-energy region for CRESST is $\lesssim$40 keV and QF$\gtrsim$10, the energy detected as light from nuclear recoils is $<$40 eV (i.e., much below 170 eV).
It can therefore be expected that an amplification of the light signal as provided by the Neganov-Luke effect improves both the energy threshold as well as the energy resolution for nuclear recoils.

II) \textit{Neganov-Luke amplification.} To verify this expectation we used this detector for a new set of measurements even under more difficult conditions due to the presence of considerable electronic noise, e.g., concerning applications at an accelerator beamline \cite{Ciemniak}. In fact, the energy threshold without applying a Neganov-Luke voltage increased to 5$\sigma\sim$185 eV.
When applying a Neganov-Luke voltage to a light detector based on this design the signal is distorted due to the occurrence of additional noise, which is mainly contained in the frequency region below 10 kHz \cite{Isaila}. This, however, is exactly the frequency region characterizing the phonon signals. It is therefore very likely that the additional noise originates from loosely bound electrons and holes that escape from their shallow trapping sites when applying a voltage to the silicon substrate. In this way, they can be drifted by the applied voltage inducing a small phonon signal corresponding to the observed noise. Since the number of these trapping sites is finite the level of this additional noise decreases with time. Therefore, in order to erode the shallow traps present in the silicon substrate more efficiently a voltage higher by 30 V than the nominal voltage intended to be used is first applied to the detector for $\sim$30 min and thereafter reduced to the nominal voltage \cite{Isaila}. This procedure heavily suppresses the noise level in the low-frequency region up to nominal voltages of $\sim$150 V. In particular, the excess noise first discussed in Ref. \cite{Stark} can be tremendously reduced.\footnote{At still higher voltages (e.g., 200 V), additional shot-noise like distortions lead to a substantially increased noise level. This noise does not change significantly with time. Therefore, it has to be of a different origin than the additional noise observed at low voltages and will not be considered here.}

\begin{figure}[htbp]
\begin{minipage}{0.43\textwidth}
\includegraphics[width=\textwidth]{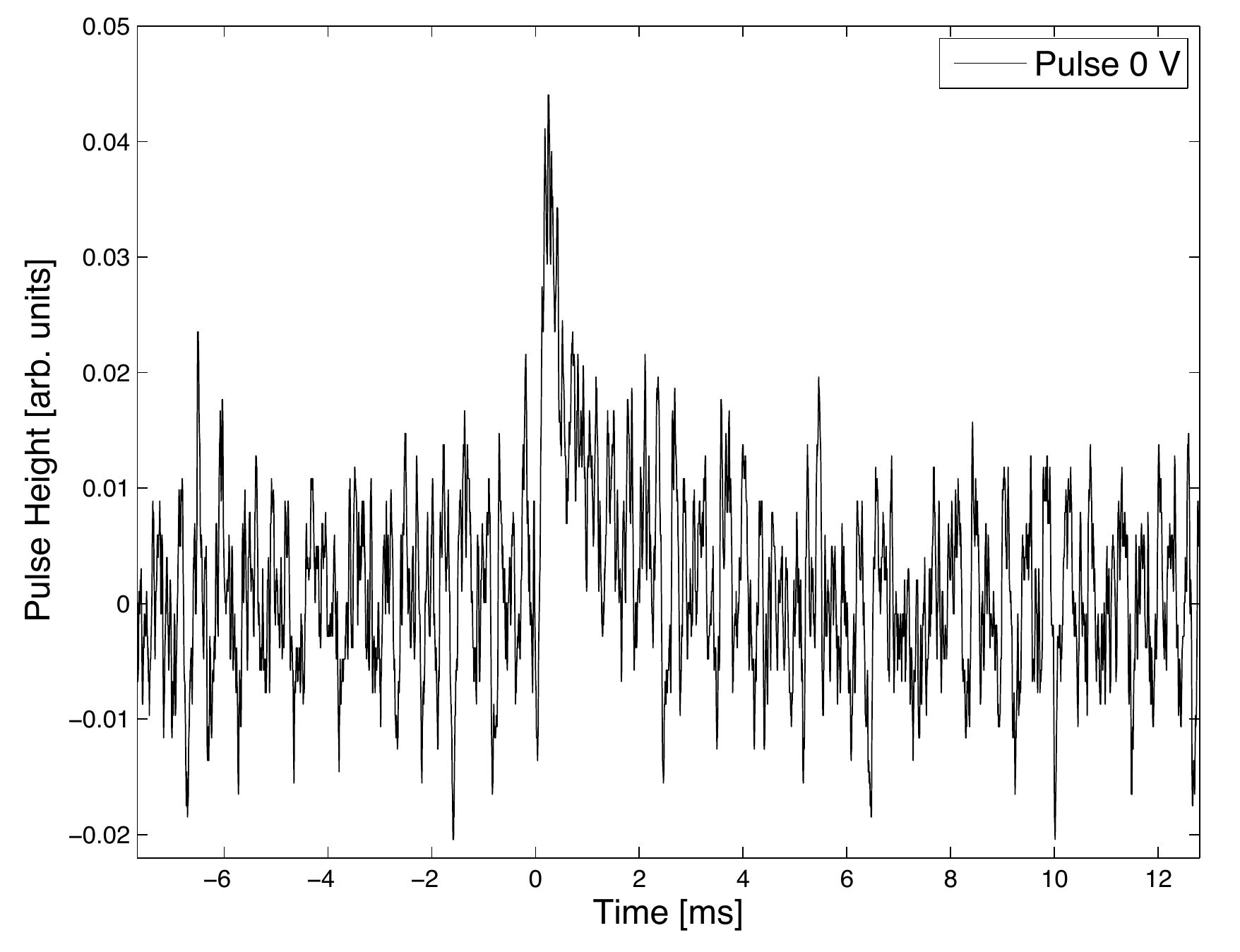}
\includegraphics[width=\textwidth]{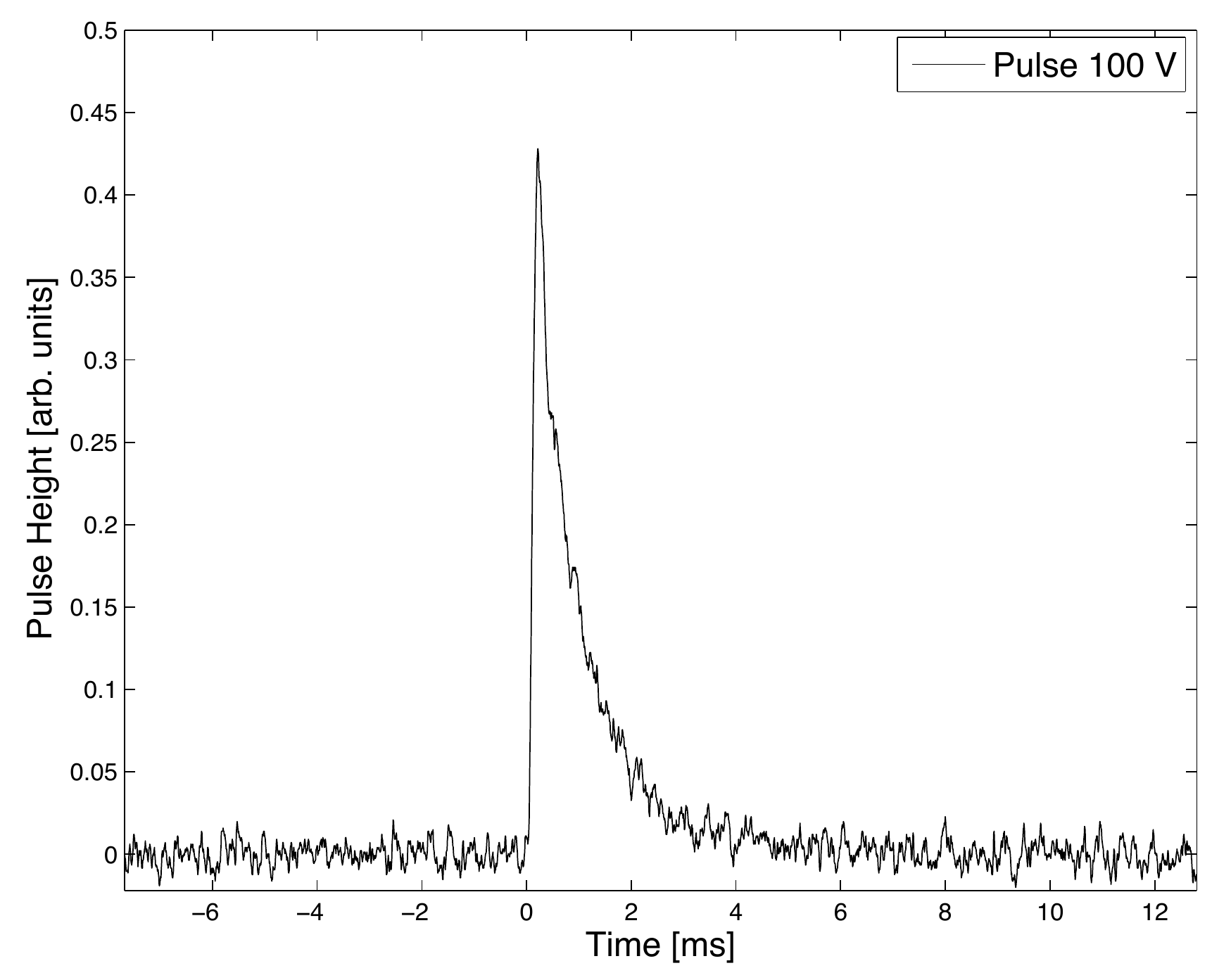}
\end{minipage}
\caption{\small{\textit{Top Panel:} Light pulse injected by a pulsed LED. The light detector is operated with no voltage applied. \textit{Bottom Panel:} Corresponding pulse with a Neganov-Luke voltage of 100 V applied. An amplification by a factor of $\sim$12 is visible. The noise level, however, is only slightly increased.}}
\label{pulses}
\end{figure}

A typical example is depicted in Fig. \ref{pulses}. The top panel exhibits a light pulse with no voltage applied. The bottom panel shows the amplified signal for a light pulse of identical intensity at a Neganov-Luke voltage of 100 V. Amplification by a factor of $\sim$12 is achieved with an improvement of the signal-to-noise ratio (S/N) by a factor of $\sim$9.


\begin{figure}[htbp]
\begin{center}
\includegraphics[width=0.63\textwidth]{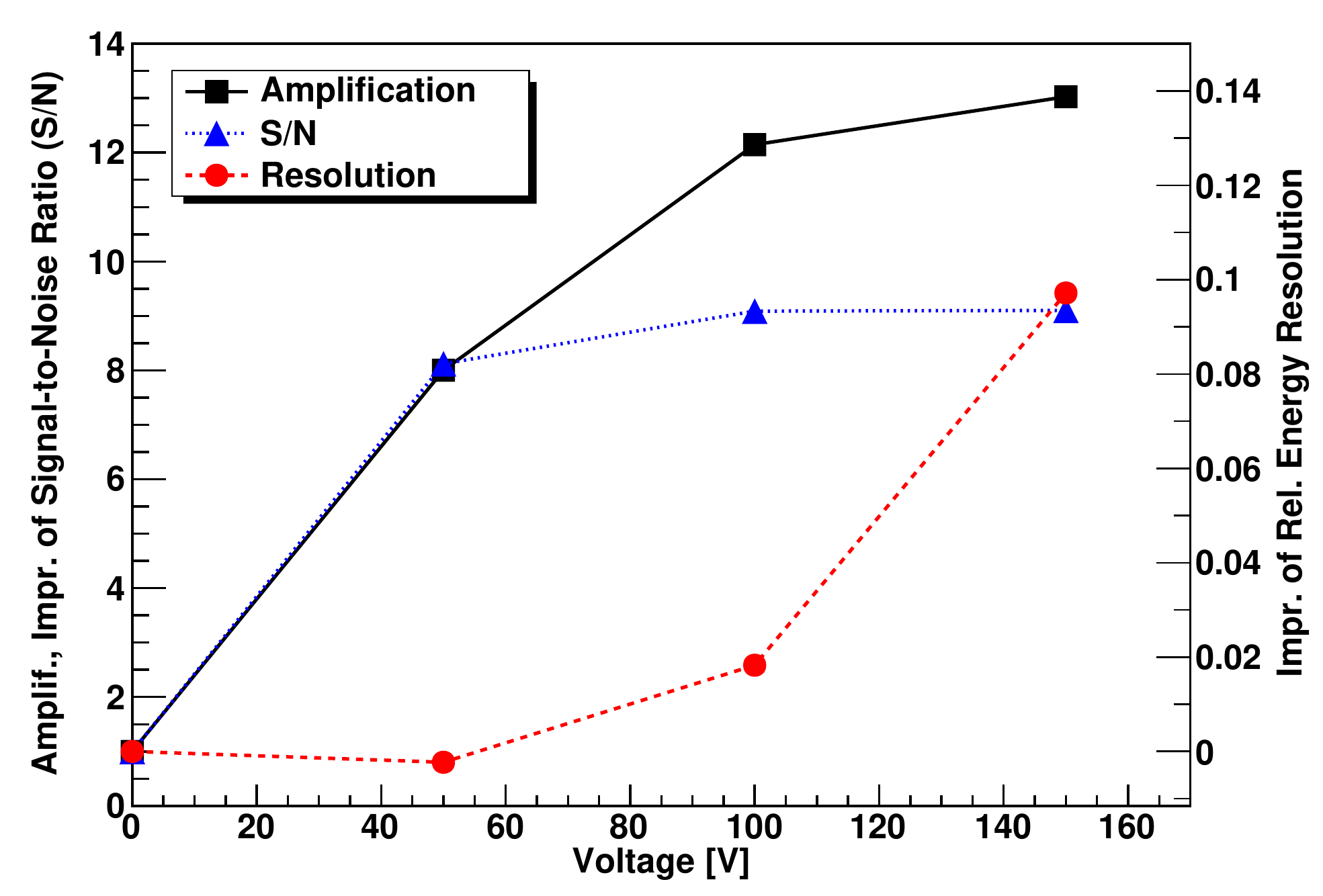}
\caption{\small{Amplification factor and improvement factor of the signal-to-noise ratio (S/N) and of the improvement of the energy resolution relative to 0 V for voltages in the range between 0 and 150 V applied by the procedure described in the main text. At 100 V an improvement of S/N by a factor of $\sim$9 is achieved. The improvement of the energy resolution for light pulses of $\sim$615 eV amounts to $\sim$10$\%$ at 150 V. The (statistical) error bars are smaller than the symbols used.}}
\label{gainsn}
\end{center}
\end{figure}

Fig. \ref{gainsn} shows the amplification factor, as well as the improvement factor (left ordinate) of S/N and the improvement of the relative energy resolution (right ordinate) compared to 0 V for voltages in the range between 0 and 150 V applied by the method described above. At voltages higher than 50 V the improvement factor of S/N is less than the amplification factor due to additional noise. At still higher voltages a saturation of both the amplification and the S/N is observed. This saturation behaviour might be explained in terms of higher trap densities created by eroding the shallow traps at high voltages. The number of drifting charge carriers is then reduced by these additional traps. An optimal performance is found at a Neganov-Luke voltage in the range of $\sim$150 V. As mentioned above (see footnote 2), at still higher voltages additional shot-noise like distortions appear.

At 150 V, an amplification factor of $\sim13$ is reached. According to eq. (\ref{Neganov-Luke}), an amplification by a factor of $\sim34$ is expected. There are several reasons why the theoretical value has not been reached \cite{Isaila}, the main reason being that the charge carriers might be trapped on their way to the electrodes before traversing the full potential V. The improvement of the S/N by a factor of $\sim$9 corresponds to an energy threshold of $E_{th}\approx$(185/9) eV $\approx$ 21 eV.

Concerning the energy resolution, our data (see Fig. \ref{gainsn}) show that with a Neganov-Luke voltage of 150 V an improvement by $\sim10\%$ at an energy of $\sim$615 eV (corresponding to a photon number of $N$=615/2.9$\approx$212) is obtained. It can be expected that at light energies $<$40 eV (caused by nuclear recoils) where $\sigma_{tot}$ is mainly determined by electronic noise an even larger improvement of the energy resolution can be reached when a Neganov-Luke voltage is applied. A lower limit is given by the energy threshold of $\sim$21 eV mentioned above. These aspects, in particular, the voltage dependence of the energy resolution (see Fig. \ref{gainsn}) will be investigated in future measurements.

\begin{figure}[htbp]
\begin{center}
\includegraphics[width=0.63\textwidth]{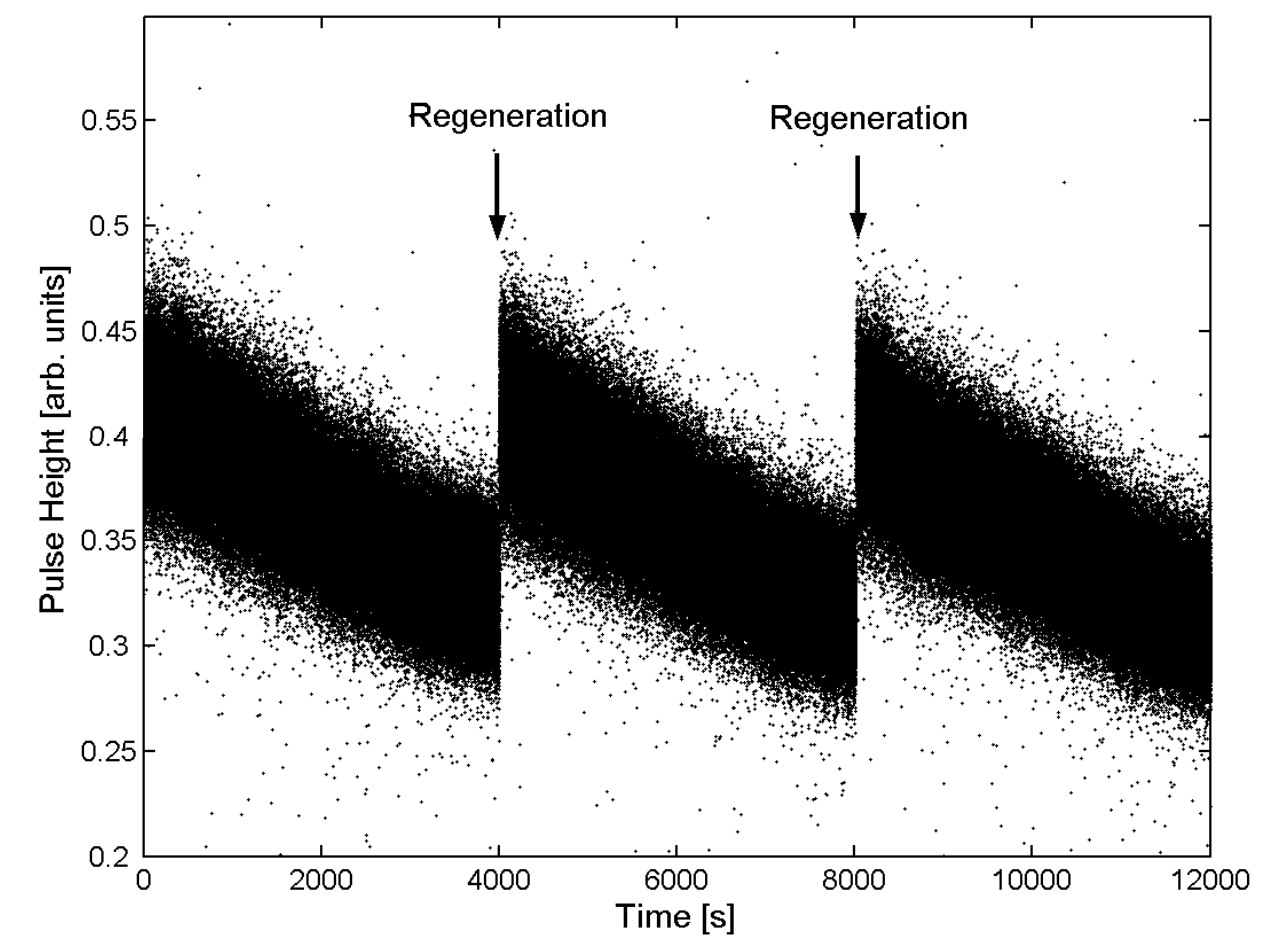}
\caption{\small{Regeneration cycles of a low-temperature light detector with Neganov-Luke amplification. Space charges can be removed by switching off the voltage and by flushing the detector for $\sim$10 s with light from the LED. Here, two regeneration cycles are depicted.}}
\label{regeneration}
\end{center}
\end{figure}

The low-temperature light detectors based on this design suffer from space charges that build up with time at the Al electrodes compensating the applied voltage which in turn leads to a decreasing pulse height with time. To remove these space charges (regeneration), the voltage is turned off and the detector is flushed with the light from a LED. Light pulses of 5 ms length (as compared to 500 ns for the calibration pulses) and a repetition rate of 50 Hz were used. In this way, the space charges can be removed within $\sim$10 s reestablishing the initial amplification. A possible temperature rise of the detector due to the flushing procedure is neutralized within a few seconds after switching off the LED. Fig. \ref{regeneration} shows the results obtained with this regeneration procedure applied two times during a running time of $\sim$3.3 h demonstrating the effectiveness of this method. Clearly, the build-up of the space charges depends on the count rate. However, due to the low count rates ($\sim$1 Hz) in deep underground laboratories, e.g., for the CRESST experiment in the Gran Sasso mountain, at most one regeneration process for every 24 h is estimated to be necessary. Therefore, the interruptions of data taking associated with the regeneration of the light detectors and the time for applying the Neganov-Luke voltage will not be significant. In addition, the reduction in pulse height with time can be monitored and taken into account using a reference light source (LED) emitting short light pulses which can easily be discriminated from the relatively long light pulses originating from the scintillating CaWO$_4$ crystal \cite{Isaila}.

In conclusion, we have shown that using the Neganov-Luke effect the signal-to-noise ratio of low-temperature light detectors can be increased by a factor of $\sim$9 leading to a drastically improved energy threshold of $\sim$21 eV. Excess noise observed with previous designs of cryogenic Neganov-Luke type detectors has been avoided. Pulses from a light-emitting diode (LED) can be used for the energy calibration. Space charges that build up at the metal electrodes and lead to a reduction of signal amplification  with time can be removed within a few seconds by flushing the detector with the light from the LED. As an example for the enormous relevance of such detectors, a possible application for the dark-matter search with the CRESST and EURECA experiments has been pointed out.

\begin{acknowledgments}
This work was supported by funds of the Deutsche For\-schungsgemeinschaft DFG (Trans\-regio 27: Neutrinos and Beyond), the Munich Cluster of Excellence (Origin and Structure of the Universe), and the  Maier-Leibnitz-Laboratorium (Garching).
\end{acknowledgments}


\begin{thebibliography}{99}

\bibitem{Taoso}
M. Taoso, G. Bertone, and A. Masiero, JCAP {\bf 03}, 22 (2008); arXiv: 0711.4996.

\bibitem{cresst}
G. Angloher {\sl et al.}, Astropart. Phys. {\bf 31}, 270 (2009).

\bibitem{cresst1}
G. Angloher {\sl et al.}, Eur. Phys. J. {\bf 72}, 1971 (2012); arXiv: 1109.0702.

\bibitem{EURECA}
H. Kraus {\sl et al.}, J. Phys.: Conf. Ser. {\bf 39}, 139 (2006).

\bibitem{Proebst}
F. Pr\"obst  {\sl et al.}, J. Low Temp. Phys. {\bf 100}, 69 (1995).

\bibitem{Bavykina}
I. Bavykina {\sl et al.}, Astropart. Phys. {\bf 28}, 489 (2007); arXiv: 0707.0766.

\bibitem{neganov}
B. Neganov and V. Trofimov, Otkrytia i izobretenia {\bf 146}, 215 (1985). 

\bibitem{luke}
P.N. Luke, J. Appl. Phys. {\bf 64}, 6858 (1988).

\bibitem{Akerib}
D.S. Akerib {\sl et al.}, Nucl. Instr. Meth. Phys. Res. {\bf A}520, 163 (2004).

\bibitem{Spooner}
N.J.C. Spooner, G.J. Homer, and P.F. Smith, Phys. Lett. B {\bf 278}, 382 (1992).

\bibitem{Stark}
M. Stark {\sl et al.}, Nucl. Instr. Meth. Phys. Res. {\bf A}545, 738 (2005).

\bibitem{Isaila}
C. Isaila, PhD Thesis, Technische Universit\"at M\"unchen, 2010.\\ http://nbn-resolving.de/urn/resolver.pl?urn:nbn:de:bvb:91-diss-20100610-980371-1-2

\bibitem{IsailaNIM} C. Isaila {\sl et al.}, Nucl. Instr. Meth. Phys. Res. {\bf A}559, 399 (2006).

\bibitem{Roth}
S. Roth {\sl et al.}, Optical Materials  {\bf 31}, 1415 (2009).

\bibitem{Nagel}
U. Nagel {\sl et al.}, J. Appl. Phys. {\bf 76}, 4262 (1994).

\bibitem{Mikhailik}
V.B. Mikhailik {\sl et al.}, Phys. Rev. B{\bf 69}, 205110 (2004).

\bibitem{Ciemniak}
C. Ciemniak, PhD Thesis, Technische Universit\"at M\"un\-chen, 2011.\\ http://nbn-resolving.de/urn/resolver.pl?urn:nbn:de:bvb:91-diss-20110706-1078311-1-5

 

\end{thebibliography}

\end{document}